\documentclass{iau} 
\usepackage{graphicx}

\title[Radionuclides in SN 1987A and Cassiopeia A] 
{Spatial distribution of radionuclides in\\ 3D models of SN 1987A and Cas~A}

\author[Hans-Thomas Janka, Michael Gabler \& Annop Wongwathanarat]   
{Hans-Thomas Janka$^1$,
Michael Gabler$^2$
\and Annop Wongwathanarat$^{1,3}$}

\affiliation{$^1$Max Planck Institute for Astrophysics, \\ 
Postfach 1317, D-85741 Garching, Germany \\ 
email: {\tt thj@mpa-garching.mpg.de}  \\[\affilskip]
$^2$Max Planck Institute for Astrophysics, \\ 
Postfach 1317, D-85741 Garching, Germany \\ 
email: {\tt miga@mpa-garching.mpg.de}  \\[\affilskip]
$^3$RIKEN, Astrophysical Big Bang Laboratory,\\ 
2-1 Hirosawa, Wako, Saitama 351-0198, Japan \\
email: {\tt annop.wongwathanarat@riken.jp}}

\pubyear{2017}
\volume{331}  
\jname{SN 1987A, 30 Years Later}
\editors{M. Renaud, A. Marcowith, G. Dubner, A. Ray \& A. Bykov, eds.}
\begin{document}

\maketitle

\begin{abstract}
Fostered by the possibilities of multi-dimensional computational modeling,
in particular the advent of three-dimensional (3D) simulations, our understanding
of the neutrino-driven explosion mechanism of core-collapse supernovae (SNe) 
has experienced remarkable progress over the past decade. 
First self-consistent, first-principle models have shown successful explosions
in 3D, and even failed cases may be cured by moderate changes of the microphysics
inside the neutron star (NS), better grid resolution, or more detailed progenitor 
conditions at the onset of core collapse, in particular large-scale perturbations
in the convective Si and O burning shells. 3D simulations have also achieved
to follow neutrino-driven explosions continuously from the initiation of the
blast wave, through the shock breakout from the progenitor surface, into the 
radioactively powered evolution of the SN, and towards the free expansion 
phase of the emerging remnant. Here we present results from such simulations,
which form the basis for direct comparisons with observations of SNe and SN
remnants in order to derive constraints on the still disputed explosion mechanism.
It is shown that predictions based on hydrodynamic instabilities and mixing
processes associated with neutrino-driven explosions yield good agreement with
measured NS kicks, light-curve properties of SN~1987A and asymmetries of 
iron and $^{44}$Ti distributions observed in SN~1987A and Cassiopeia~A.
\end{abstract}

\firstsection 
\section{Introduction}

Since the seminal works by \cite{Colgate66} and 
\cite{Arnett67} more than 50 years ago,
the computational modeling of supernova (SN) explosions has
experienced enormous progress with respect to the numerical methods,
input physics, computational accuracy, and dimensionality.
Nevertheless, despite many generations of successively improved
simulations, initially performed in spherical symmetry (1D),
then since the 1990's also in two dimensions (2D), and in recent
years also 
in full 3D, the physical processes that cause the explosion are
not established yet (for recent reviews, see 
\cite{Janka12,Burrows13,Foglizzo15,Janka16,Mueller16,Janka17b}).
For the far majority of normal
core-collapse SNe the delayed neutrino-driven mechanism
(\cite{Wilson85,Bethe85}) is widely considered
as most likely explanation, because it taps the vast
reservoir of energy that is radiated by the nascent neutron star
(NS) in neutrinos and outranges the explosion energy of SNe
by a factor of several hundred. 

The neutrino-driven mechanism
also satisfies a number of fundamental requirements that any 
viable scenario should fulfill (for a detailed discussion,
see \cite{Janka17b}). First, the mechanism is {\em not
``robust''}, because it should allow for the formation of 
stellar-mass black holes (BHs), whose abundent existence has been
confirmed by the recent measurements of gravitional waves from
binary BH mergers (\cite{Abbott16}). 
Second, it must be {\em inefficient}, because
it should explain why SN explosion energies are so much lower
than the gigantic amount of gravitational binding energy that is
available during NS or 
BH formation. Third, it should be {\em self-regulated}, because
the energy transferred to the ejecta does not largely exceed
the binding energy of the progenitor shells outside of the 
degenerate core, i.e., it is as low as several $10^{49}$\,erg to about
$10^{50}$\,erg near the low-mass side of SN progenitors and may be
$\sim$$(1-2)\times 10^{51}$\,erg for energetic explosions of stars
around 20\,$M_\odot$. This clearly separates such normal SNe
from the significantly more powerful but much rarer hypernovae
(with a rate of less than roughly one out of thousand 
core-collapse events), whose energies and
explosion properties point to another mechanism, probably 
invoking the formation of BHs or magnetars and of jet-driven 
outflows caused by extreme amplification of magnetic fields
during the collapse of rapidly rotating progenitors 
(see, e.g., \cite{Woosley06}). The latter are the 
final outcome of very special and uncommon
single and binary star evolution scenarios of massive stars
(e.g., \cite{Levan16}).

Although first 3D simulations with energy-dependent neutrino
transport have meanwhile obtained successful explosions by 
neutrino heating 
(\cite{Takiwaki14,Melson15a,Melson15b,Lentz15,Roberts16,Mueller16}),
the viability of this theoretical scenario is still not
generally accepted (e.g., \cite{Soker17a,Soker17b,Kushnir15,Blum16}).
Doubts are either motivated by refering to those computational models
where the numerical setups still failed to produce explosions,
or they are justified by pointing to remaining shortcomings of the
current calculations, for example the lack of a clear demonstration
by modern simulations that neutrino-driven explosions can yield 
energies around $10^{51}$\,erg or more (e.g., \cite{Soker17a} and 
references therein). Such missing pieces in the puzzle are very
likely to fall into place once longer and better resolved 3D 
simulations are performed, larger sets of progenitors are
investigated, more consistent initial 
conditions that account for large-scale perturbations in
the convective silicon and oxygen burning shells are applied
(\cite{Couch13,Couch15,Mueller15,Mueller16}),
and further improvements of the microphysics are
used for the description of dense NS matter and of neutrino
interactions in the correlated nuclear medium
(see the sensitivity test by \cite{Melson15b}).

Moreover, convincing alternatives to the neutrino-heating
mechanism, which could elucidate the
processes that initiate and power most core-collapse SNe,
do not exist. Considering thermonuclear burning in the stellar
carbon shell during gravitational collapse as the main energy
source of the SN blast wave (\cite{Kushnir15}) demands a
radical change of the chemical structure of progenitor stars
in conflict with the results of stellar evolution calculations.
This approach to overrule the common notion of stellar evolution
appears as an unnecessary and overmotivated act of desperation. 
Similarly unsatisfactory (and not particularly elegant)
is a renunciation of self-consistency in the explosion
modeling by introducing, in an ad hoc manner, components whose
physical origin remains unexplained, for example ``jittering jets''
(\cite{Soker17a} and references therein). Magnetic fields, which
are often readily invoked to bridge gaps of reasoning,
cannot be made responsible for the production of (jittering) jets 
in most or even in all SNe as claimed by \cite{Soker17a}.
Many stellar collapse simulations including the effects of
magnetic fields (e.g., 
\cite{Moiseenko07,Burrows07,Moesta14,Moesta15,Obergaulinger17})
have demonstrated that very rapid 
rotation of the progenitor cores is needed to obtain
magnetohydrodynamic jets by field amplification. The thus required
angular momentum in stellar cores is in conflict
with the moderate spin rates of white dwarfs (\cite{Kawaler15})
and young neutron stars (\cite{Heger00,Heger05}),
with predictions from stellar evolution calculations including
angular momentum transport through magnetic fields for
progenitors of normal core-collapse SNe (\cite{Heger05}),
and with astroseismological measurements, which find
core rotation rates that are considerably {\em slower} than 
predicted by the stellar evolution models
(e.g., \cite{Beck12,Eggenberger12,Eggenberger16}).
Angular momentum separation in collapsed stellar cores by spiral
modes of the standing accretion shock instability (SASI), which
can develop even in nonrotating stars (e.g.\ \cite{Kazeroni17}), 
has been shown to potentially lead to considerable growth of the
initial B-fields (\cite{Endeve12}). However, the emerging field 
configuration is highly turbulent and does not possess the ordered
structure on large scales that is necessary for driving jet outflows.
This problem will not disappear by future 3D simulations with
increased grid refinement, because better resolution of the turbulent
flow will foster a growth of the fields on {\em small scales}
while it is unlikely to boost a large-scale field that could enable
jet formation.

Despite these direct and indirect arguments in favor of 
neutrino-driven explosions, this theoretical scenario requires
further consolidation by demonstrating that corresponding
explosion models are able to account for the observational
properties of SNe and of their remnants. In the following,
a number of such aspects will be briefly summarized that 
provide support for the neutrino-driven mechanism.

\begin{figure}[!]
\begin{center}
 \includegraphics[width=5.5cm]{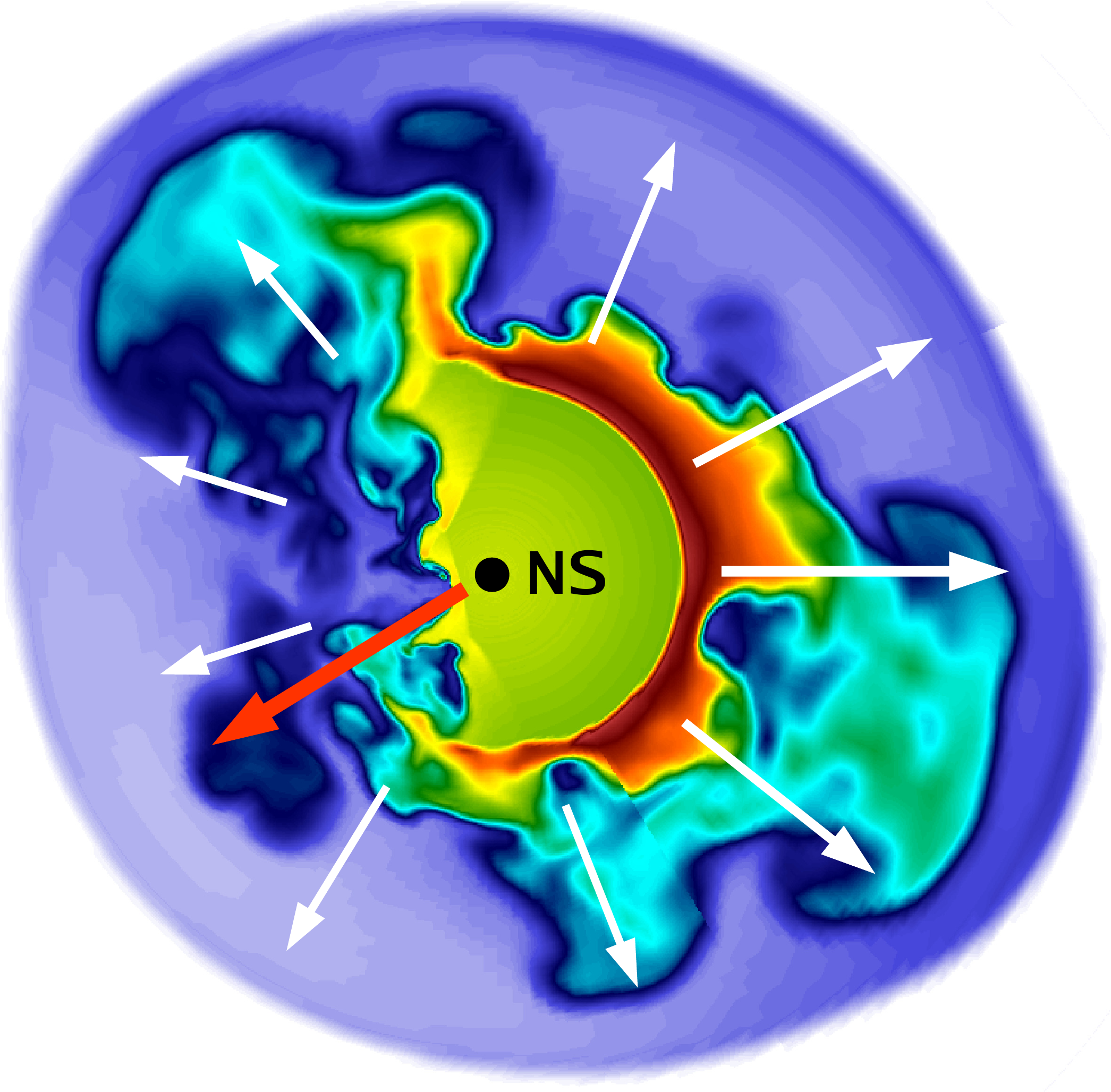}
 \includegraphics[width=7.5cm]{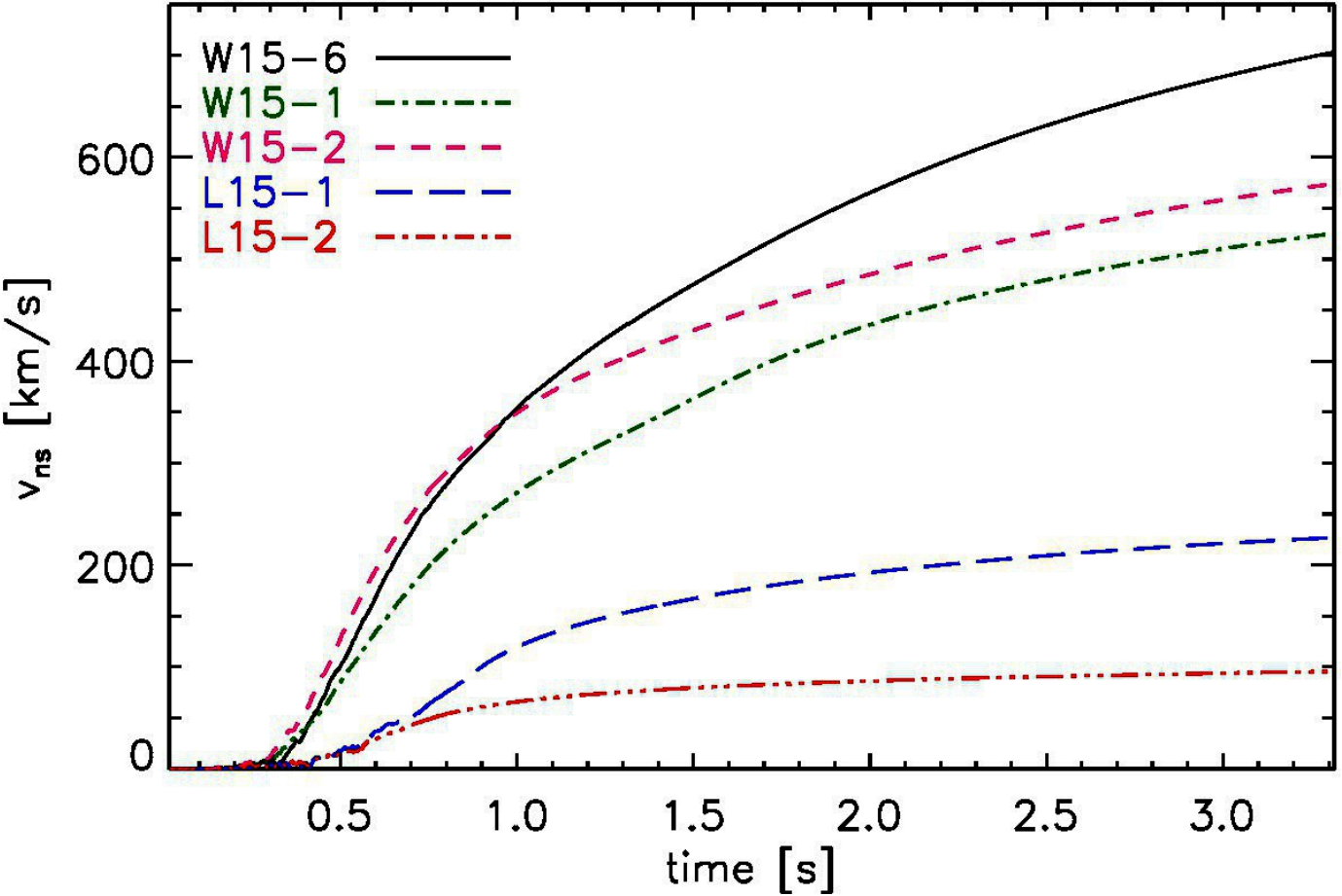}
 \caption{{\em Left:} Mass ejection asymmetry in a cross-sectional plane
for one of the 3D simulations of neutrino-driven explosions (at 1.3\,s
after core bounce) published by \cite{Wongwathanarat13}.
The NS kick direction (marked by the red arrow) is opposite to the
hemisphere of the stronger explosion, where relatively more mass from
silicon to the iron group is expelled. The entropy per nucleon
is color coded (blue, green, yellow, red signal growing values),
and the outer contour is the deformed SN shock.
{\em Right:} NS star kick velocities as functions of time for a subset
of the 3D simulations of neutrino-driven explosions in
\cite{Wongwathanarat13}.}
\label{fig:NSkick}
\end{center}
\end{figure}

\section{Neutrino-driven explosions: 3D models versus observations}

In the following we present some selected results from recent
3D calculations of neutrino-driven explosions that were performed
with the \textsc{Prometheus-HOTB} code for
different red (RSG) and blue supergiants (BSG) 
of 15\,$M_\odot$ and 20\,$M_\odot$
by \cite{Wongwathanarat13,Wongwathanarat15,Wongwathanarat16}. The blast waves
were artifically initiated with a parametric neutrino ``engine''
such that explosion energies in the ballpark of quite energetic SNe
like SN~1987A and Cassiopeia~A were obtained. Although the modeling
is not based on a fully self-consistent first-principle approach, 
the simulations still
capture the essential physics of neutrino-driven explosions, which
start by the energy deposition of neutrinos around the newly formed
NS and succeed by the crucial support of hydrodynamic instabilities
(convective overturn and SASI), creating large-scale asymmetries
in the outgoing SN shock and the innermost ejecta already during
the very first second of the explosion. The subsequent SN evolution 
is then followed continuously from core bounce to shock breakout
at the progenitor surface and beyond (see the contribution in this
volume by M.~Gabler et al.).

{\underline{\it Neutron-star kicks}}.
The asymmetric onset of the explosion causes momentum
transfer to the newly formed NS mainly by the long-time 
gravitational interaction between the compact remnant and
the anisotropically expelled matter 
(\cite{Scheck06,Wongwathanarat13}). This leads to NS kicks
opposite to the direction of the stronger explosion (consistent
with linear momentum conservation in the disrupted star; 
Fig.~\ref{fig:NSkick}, left). The kick velocities $v_\mathrm{ns}$
cover the full range of measured space velocities of young
NSs up to nearly 1000\,km\,s$^{-1}$ 
(Fig.~\ref{fig:NSkick}, right):
\begin{equation}
v_\mathrm{ns} = 211\,\frac{\mathrm{km}}{\mathrm{s}}\,\,\zeta\,
\left(\frac{\alpha_\mathrm{ej}}{0.1}\right)
\left(\frac{E_\mathrm{exp}}{10^{51}\,\mathrm{erg}}\right)
\left(\frac{M}{1.5\,M_\odot}\right)^{-1} ,
\label{eq:vns}
\end{equation}
(\cite{Janka17a}), 
where $\alpha_\mathrm{ej}$ is the momentum-asymmetry
of the innermost ejecta, $E_\mathrm{exp}$ the explosion energy,
$M$ the NS mass, and $\zeta$ is a numerical factor of order unity. 
$\alpha_\mathrm{ej}$ is determined by stochastic effects
during the onset of the asymmetric explosion. While it is found
to vary usually between 0 and $\sim$0.3 in existing 2D and 3D 
SN models, higher values are well possible, and
kick velocities in excess of 1000\,km\,s$^{-1}$ seem in
reach for extreme explosions with high energies and large 
asphericities.

\begin{figure}[b]
\begin{center}
 \includegraphics[width=3.20cm]{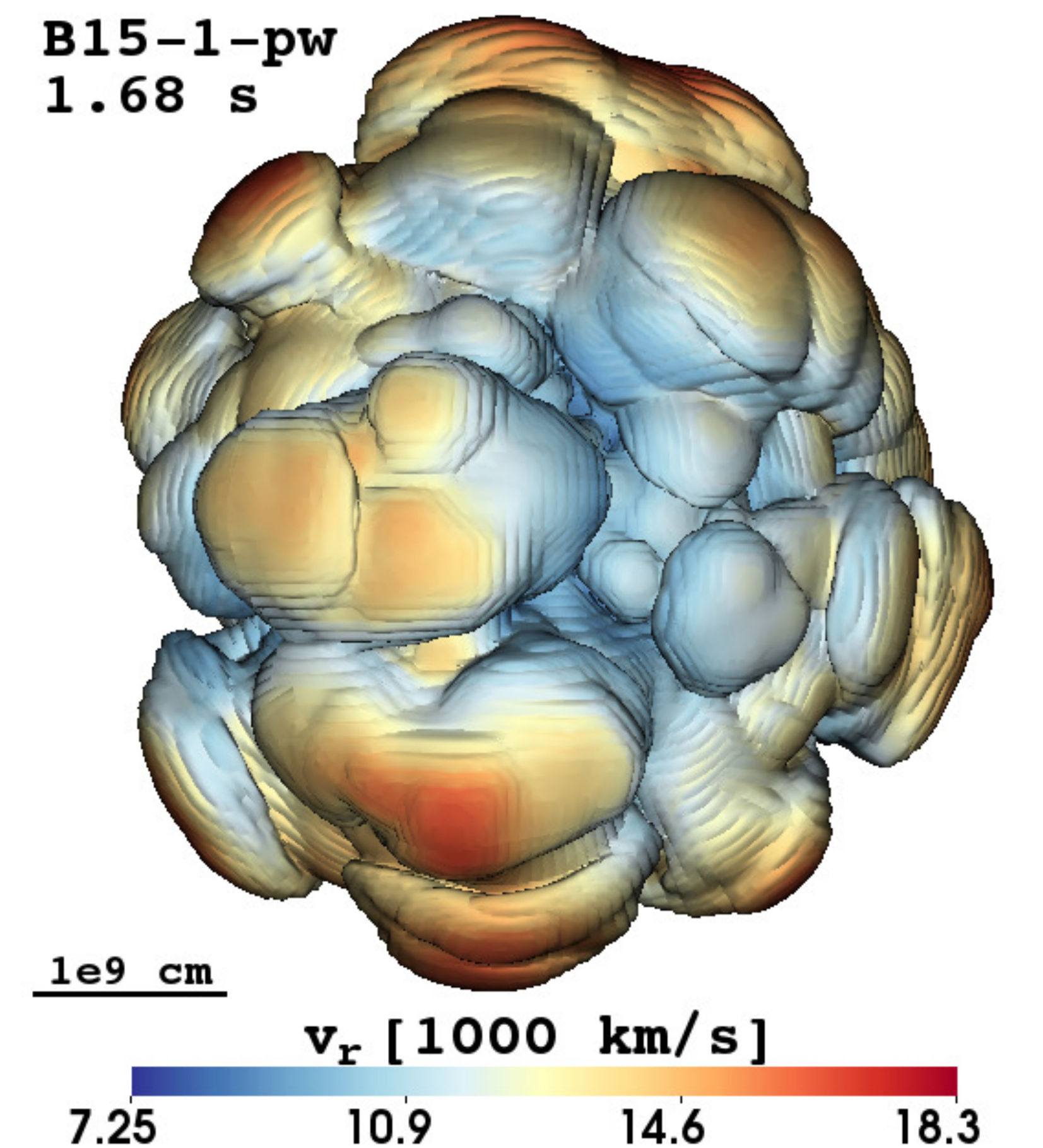}
 \includegraphics[width=3.20cm]{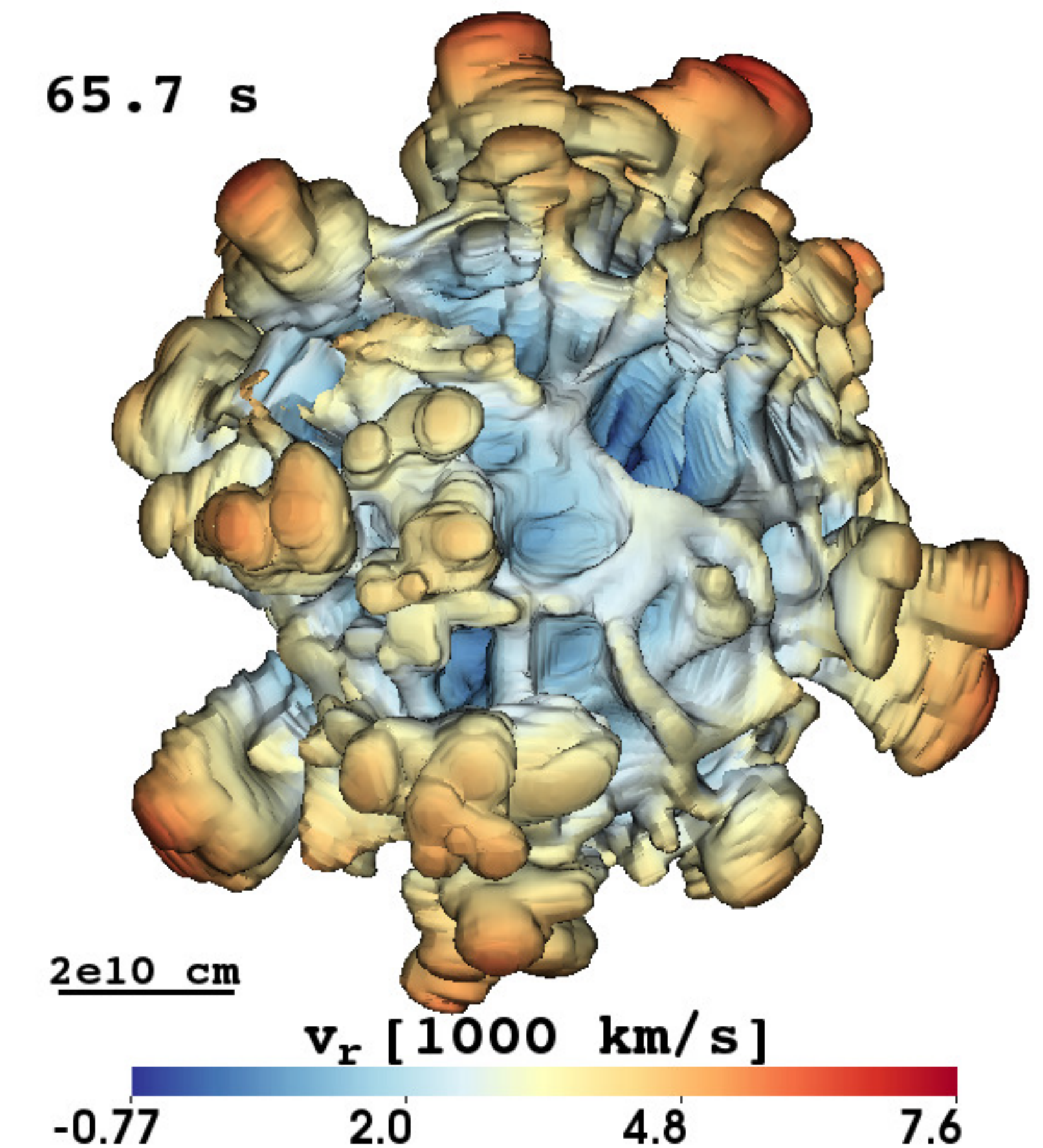}
 \includegraphics[width=3.20cm]{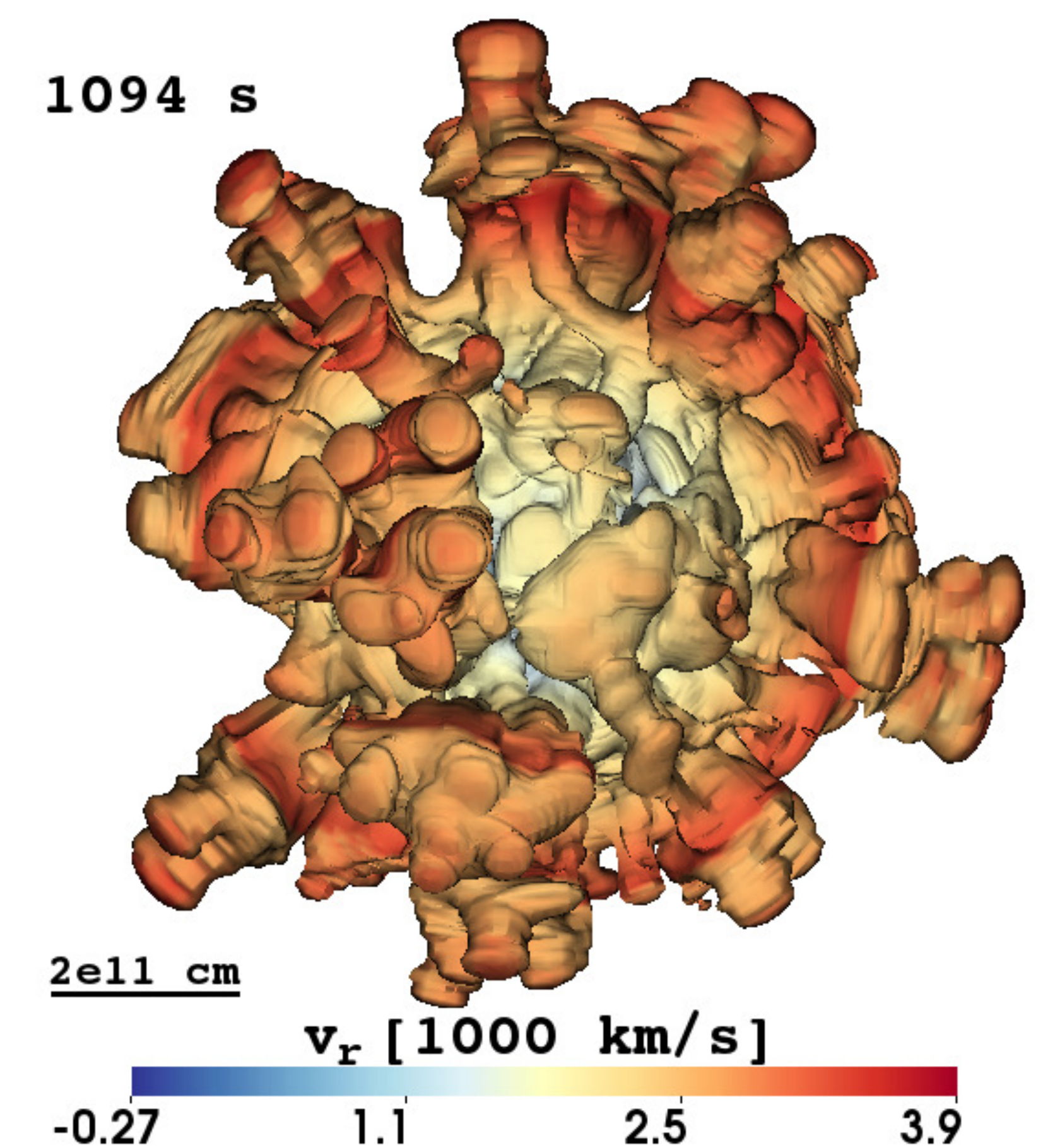}
 \includegraphics[width=3.20cm]{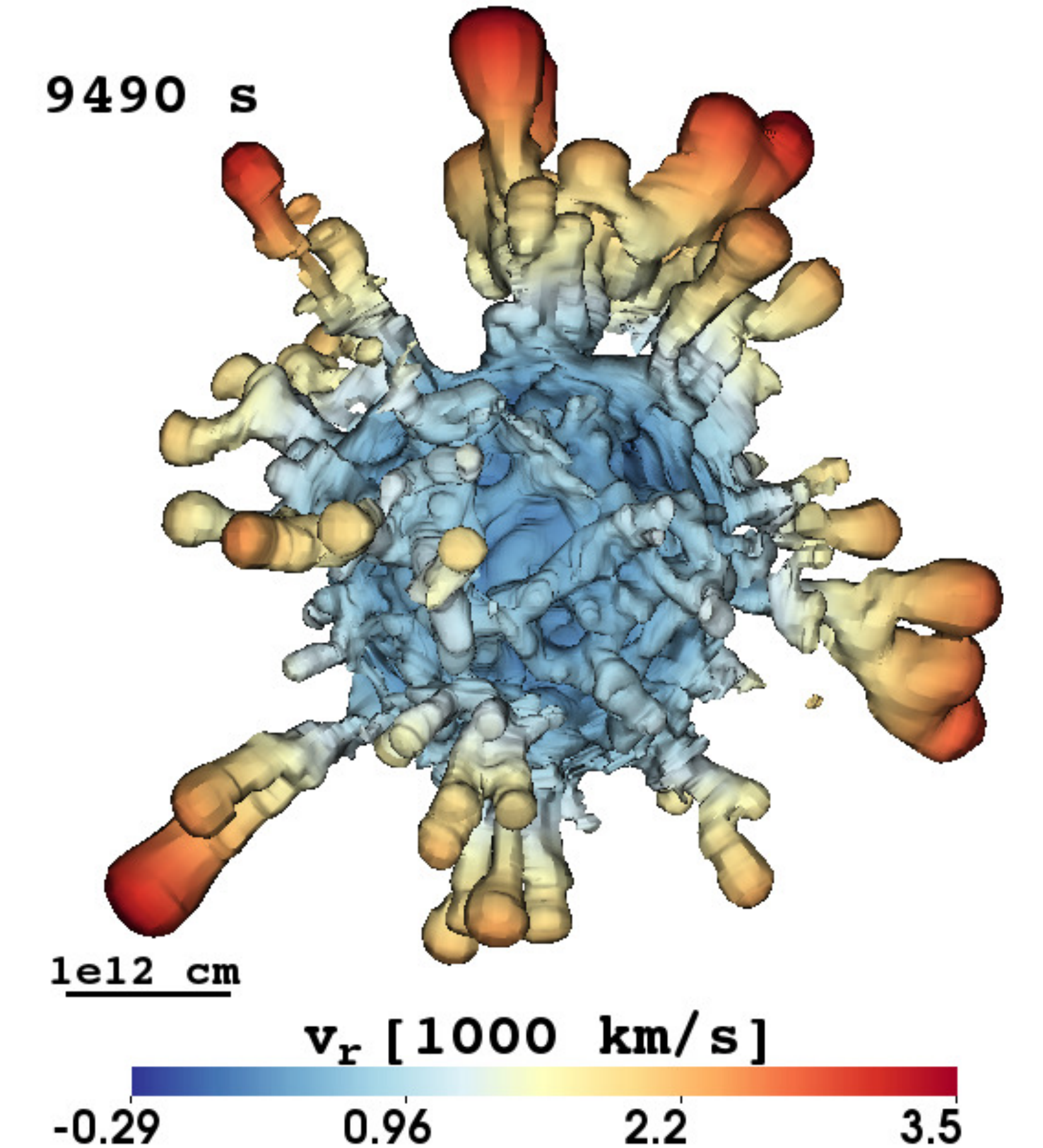}
 \caption{Fragmentation and growth of initial explosion asymmetries 
by secondary Rayleigh-Taylor instabilities at the C-O/He and He/H
composition interfaces for a 15\,$M_\odot$ BSG star. The images show
isosurfaces for a mass-fraction of 3\% iron-group elements at the
time when the SN shock crosses the C-O/He and He/H interfaces
({\em left} and {\em second}, respectively), before the reverse
shock from the He/H interface hits the iron ejecta ({\em third}),
and at shock breakout ({\em right}). (Images from 
\cite{Wongwathanarat15})}
   \label{fig:B15evol}
\end{center}
\end{figure}
\begin{figure}[!]
\begin{center}
 \includegraphics[width=9.4cm]{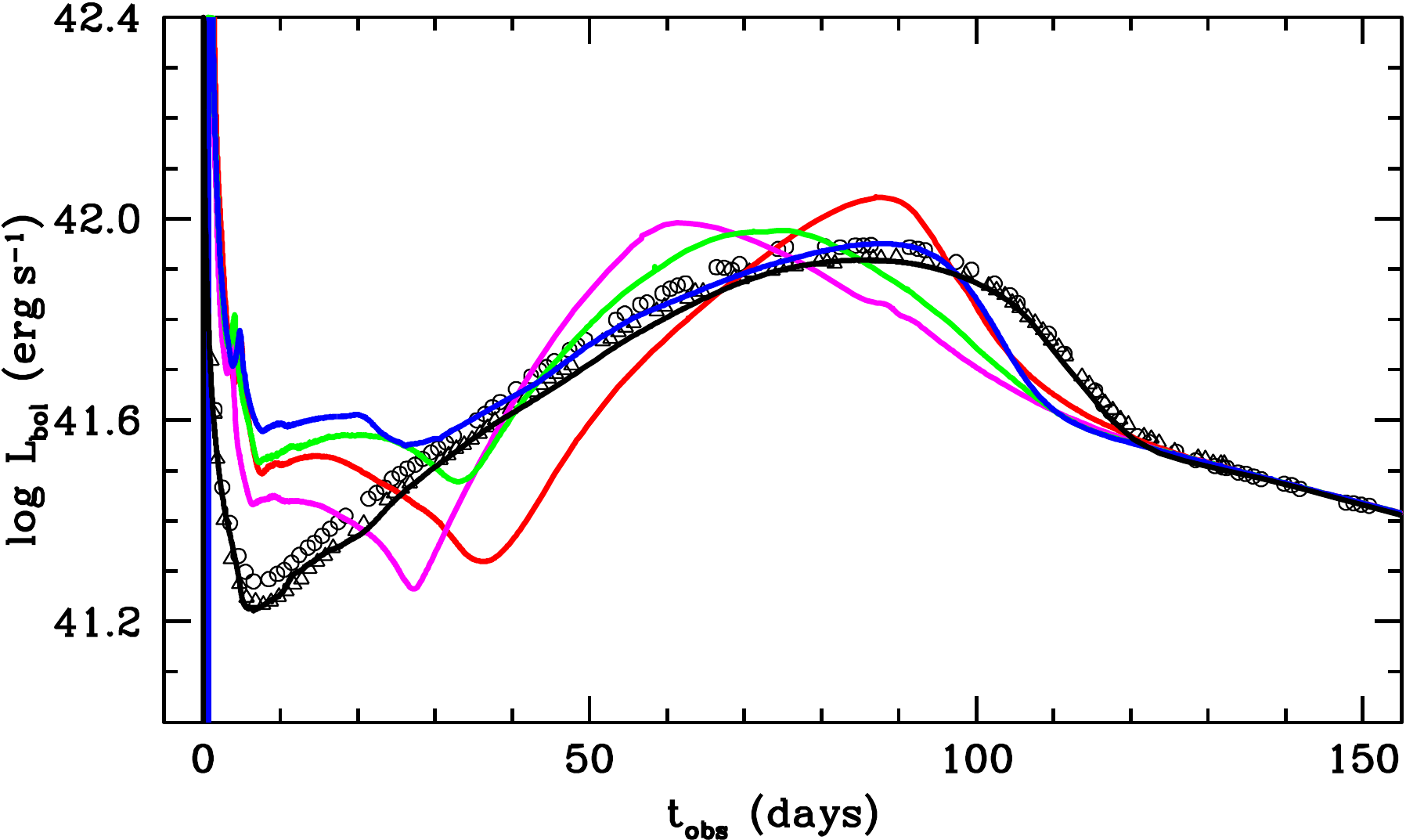}
 \caption{Bolometric light curves for SN calculations of four BSG
progenitors (solid colored lines) compared to the observed light curve 
of SN 1987A (symbols; see \cite{Utrobin15}). The neutrino-driven
explosions from shock formation until beyond shock breakout were
simulated in 3D, while the long-time radiation-hydrodynamical light-curve
calculations were carried out
in 1D, starting from spherically averaged data of the 3D explosion models.
The blue line corresponds to the model shown in Fig.~\ref{fig:B15evol},
where strong outward mixing of radioactive nickel and inward mixing of 
hydrogen allow for a good reproduction of the light-curve peak of SN~1987A.
The imperfect match around the luminosity minimum and during the decline
phase from the peak is a consequence of an overestimated radius and 
underestimated ejecta mass by the progenitor model. The black line
corresponds to a specifically tailored stellar model.}
   \label{fig:SN1987Alc}
\end{center}
\end{figure}
\begin{figure}[!]
\begin{center}
\includegraphics[width=7.4cm]{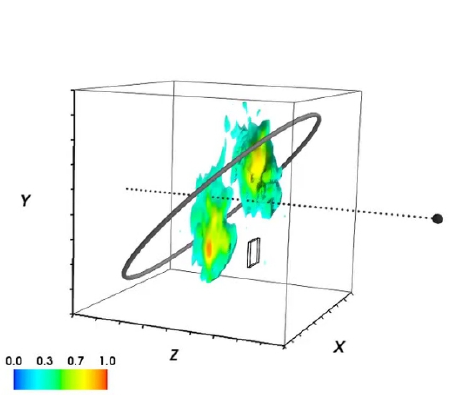}
\includegraphics[width=5.4cm]{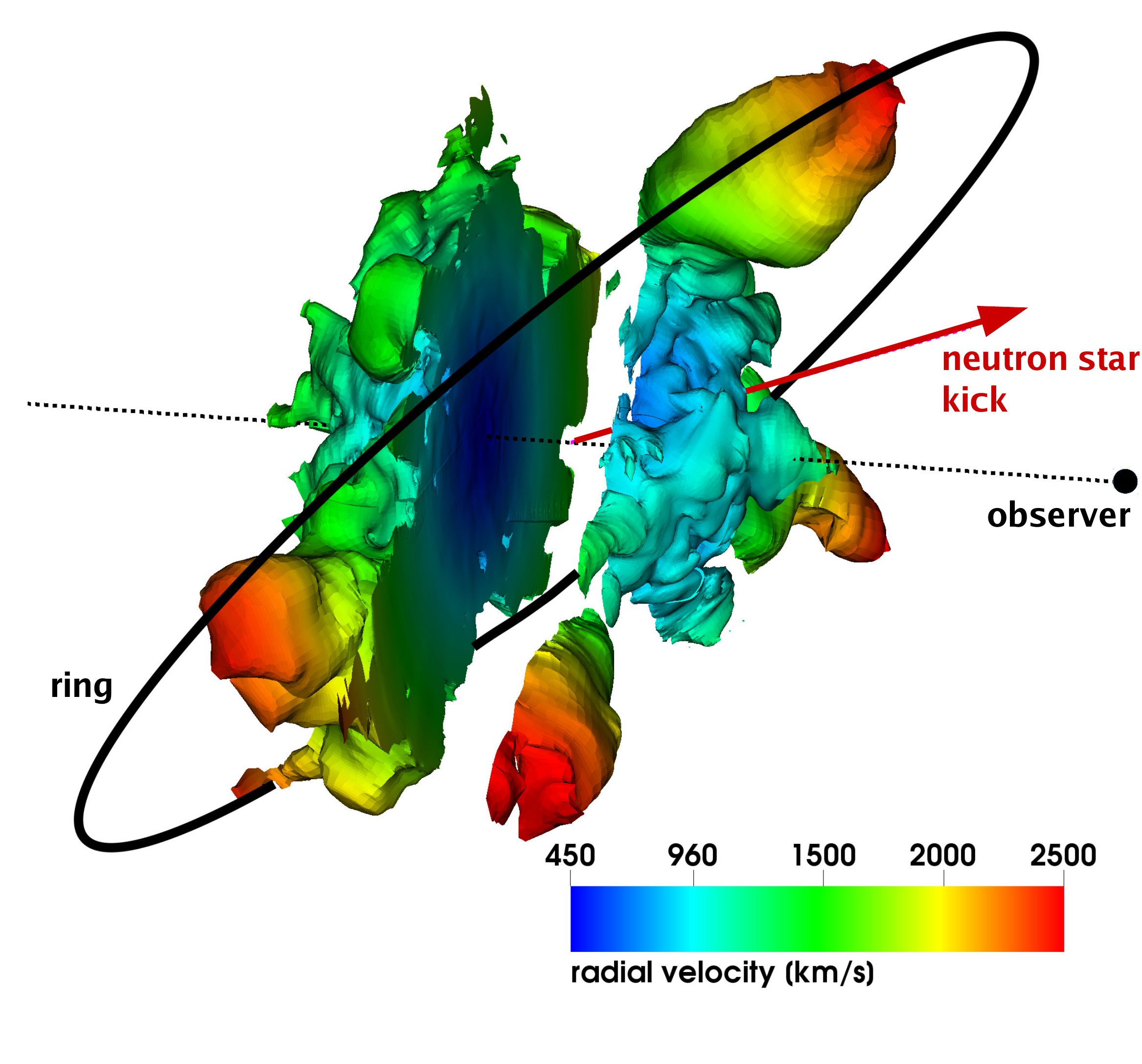}
 \caption{{\em Left:} 3D isosurfaces for [SiI]+[FeII] in SN~1987A, combining the
levels for 30\%, 50\%, and 70\% of the maximal intensity as indicated by the color
bar from \cite{Larsson16}. The ring corresponds to the position of the
reverse shock at the inner edge of the equatorial ring, and the dotted line marks
the viewing direction with the observer location indicated by the filled circle.
Tick marks on the axes are spaced in intervals of 1000\,km\,s$^{-1}$.
{\em Right:} Similar visualization of the iron distribution in a 3D
neutrino-driven explosion simulation with an energy close to
that of SN~1987A. The explosion has been evolved to 600 days
including heating by $^{56}$Ni decay. The image shows
an isosurface of constant mass-density of $^{56}$Fe with the
radial velocity color coded. Including silicon yields a very similar
morphology but higher velocities. As in the observation, the
central region within $\pm$450\,km\,s$^{-1}$ along the line of sight has been
removed. The observer direction (dashed line with bullet for observer
position) was chosen for optimal similarity with the
observations; the corresponding ring plane and the NS kick direction of the
SN model are also indicated.}
   \label{fig:SN1987Airon}
\end{center}
\end{figure}

{\underline{\it Light curve of SN 1987A}}.
The initial ejecta asymmetries, which are caused by the
neutrino-driven mechanism and its associated hydrodynamic
instabilities, lead to a global asphericity of the explosion.
They also trigger efficient growth of secondary Rayleigh-Taylor 
(RT) instablities at the C-O/He and He/H shell interfaces of
the progenitor by crossing density and pressure gradients
after the passage of the outgoing shock (for a detailed
discussion, see \cite{Wongwathanarat15}).

These RT instabilities induce the fragmentation of
the initial large-scale structures and counteract the
deceleration of the inner ejecta by the reverse shock that
moves into the metal core. As a consequence, effective outward
mixing of radioactive $^{56}$Ni (in extended finger-like 
plumes with velocities up to $\sim$3500\,km\,s$^{-1}$,
see Fig.~\ref{fig:B15evol}) and inward mixing of hydrogen
(to velocities close to zero) 
can take place. 3D simulations of neutrino-driven explosions
are thus able to account for the radial mixing and high nickel
velocities that are needed to explain the shape of the light
curve of SN~1987A (Fig.~\ref{fig:SN1987Alc}), the early 
observations of gamma-rays and X-rays from radioactive decays,
and a variety of spectral features of this SN (\cite{Arnett89}
and references therein).

{\underline{\it Silicon and iron ejecta in SN 1987A}}.
Based on detailed spectral and imaging observations with HST/STIS and 
VLT/SINFONI, \cite{Larsson16} produced 3D maps of various
chemical components of SN~1987A including silicon and iron. The morphology
of these tracers of the innermost ejecta is shown in the left panel
of Fig.~\ref{fig:SN1987Airon}. In the right panel of this figure
we display the iron distribution (Si+Fe only differs insignificantly)
in a 3D neutrino-driven explosion model for a 15\,$M_\odot$ star with
properties (energy, ejecta geometry) resembling those of SN~1987A.
We only display the dense, inner regions of the iron material up to
radial velocities of 2500\,km\,s$^{-1}$, guided by the space
velocities of the centers of the brightest regions in the left panel.
Moreover, following the observational image we also omit all mass 
with velocities between
$-$450\,km\,s$^{-1}$ and $+$450\,km\,s$^{-1}$ along the line of sight.
Assuming that 15--30\% of high-velocity Fe and Si, of which some
are in extended fingers similar to those in
Fig.~\ref{fig:SN1987Airon}, remain invisible is justified
for three reasons: First, although such high-velocity material is not 
seen in the observations, it {\em must} be present in SN~1987A to
explain its light curve and spectral properties
(see above). Second, the fingers and plumes of high-velocity
Si and Fe possess an average mass-density that is at least a
factor of 10 lower than the density of the central bulk of iron.
Third, the observed [SiI]+[FeII] emission is powered by
positrons from $^{44}$Ti decay (\cite{Larsson16}),
but titanium in the extended plumes is less abundant
(at least a factor of 2) relative to iron than in the more
concentrated, lower-velocity bulk. Therefore, the excitation of
Si and Fe by $^{44}$Ti-decay positrons must be expected to be less
efficient. The resemblance of the two images 
in Fig.~\ref{fig:SN1987Airon} is assuring, in particular since
the 3D simulation was {\em not tailored} to match SN~1987A.

\begin{figure}[!]
\begin{center}
 \includegraphics[width=6.25cm]{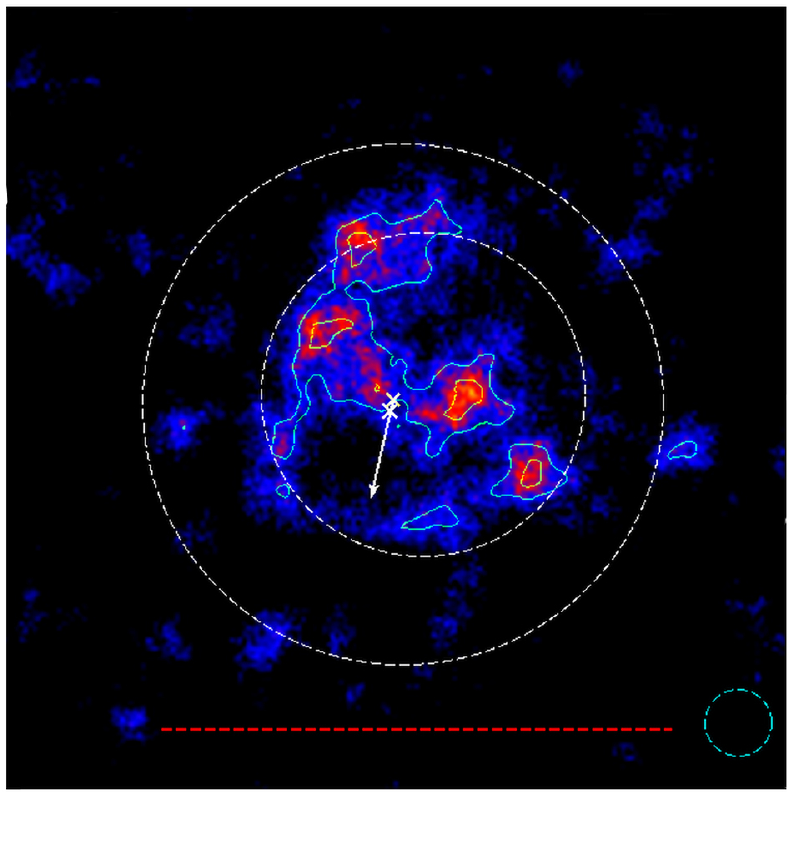}
 \includegraphics[width=6.75cm]{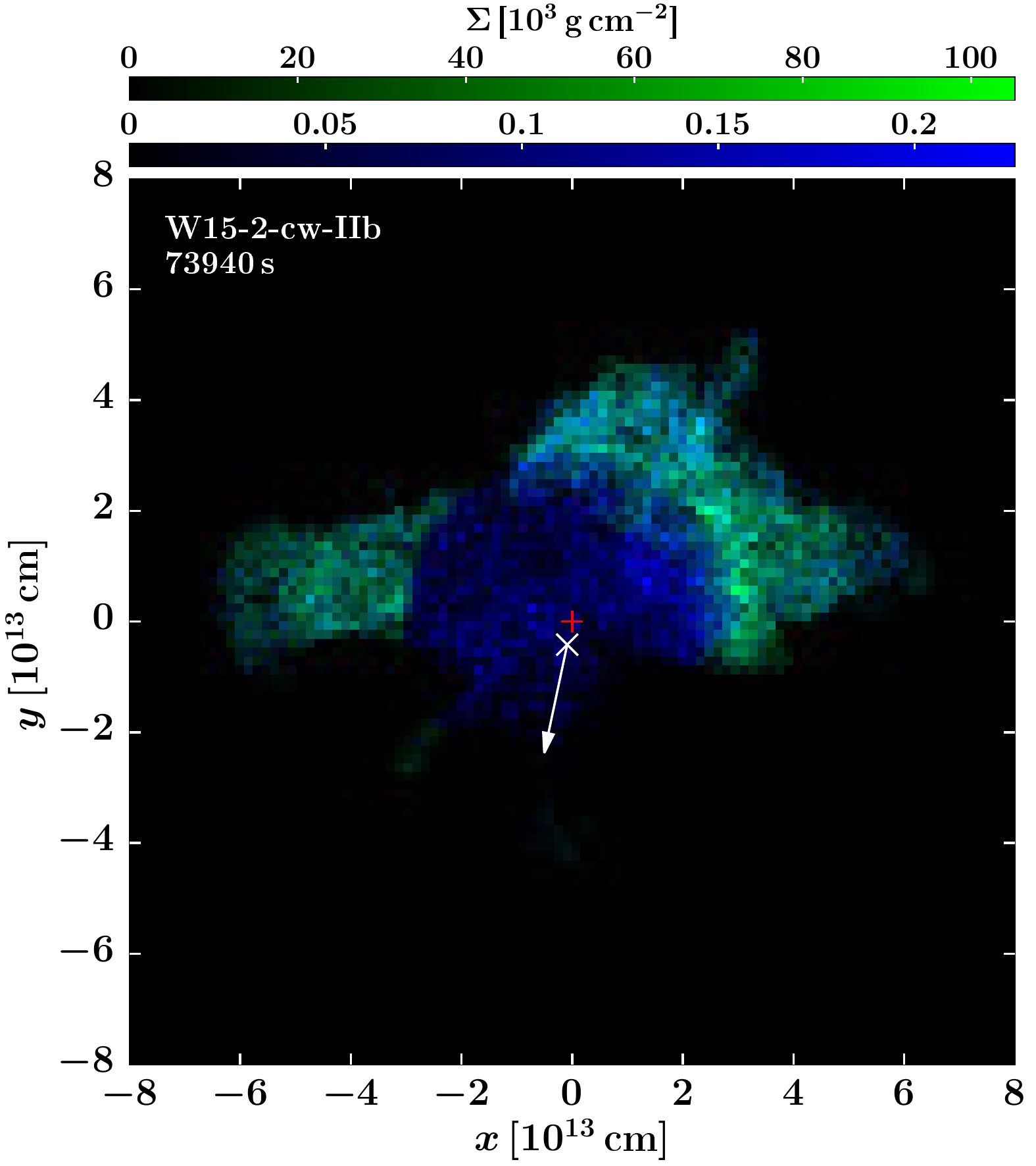}
 \caption{{\em Left:} $^{44}$Ti distribution in Cas~A with forward
and reverse shock locations marked by dashed circles
(reprinted by permission from Macmillan Publishers Ltd: Nature,
\cite{Grefenstette14}, \copyright2014).
{\em Right:} Distribution of $^{44}$Ti (blue) and $^{56}$Fe (decay
product of $^{56}$Ni) in a 3D simulation of a neutrino-driven explosion
(\cite{Wongwathanarat16}). The reverse shock is assumed to have
moved inward through half of the iron, which is therefore visible and
displayed only in the shock-heated outer shell. In both images the
geometrical center of the expansion and the location and kick
direction of the NS are indicated.}
   \label{fig:CasAiron+ti}
\end{center}
\end{figure}
\begin{figure}[!]
\begin{center}
 \includegraphics[width=6.75cm]{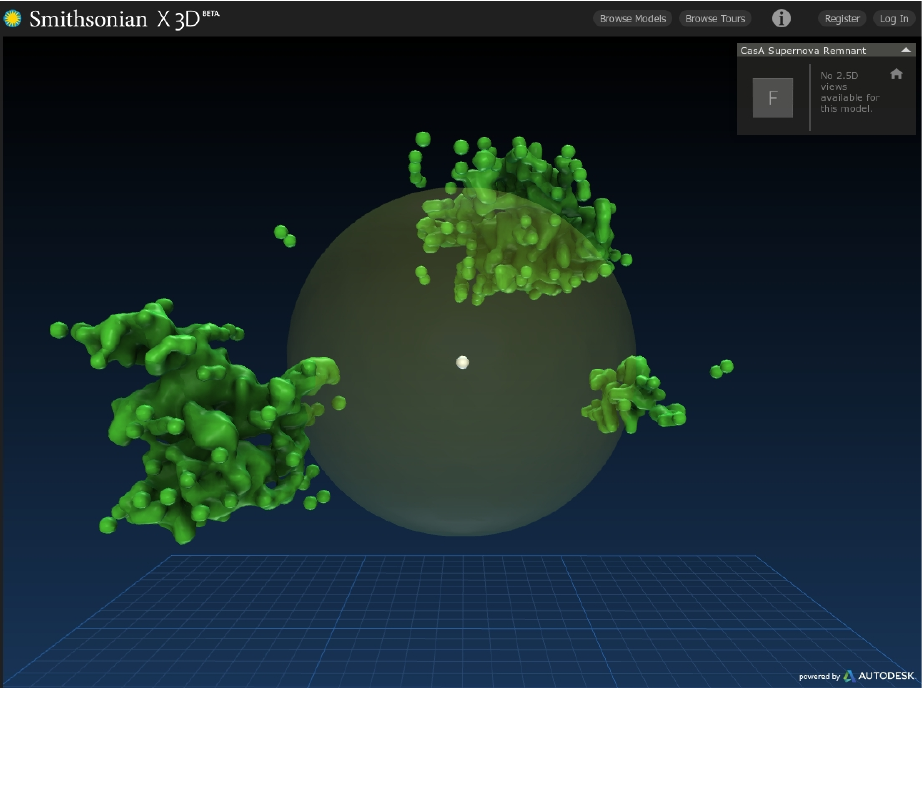}
 \includegraphics[width=6.25cm]{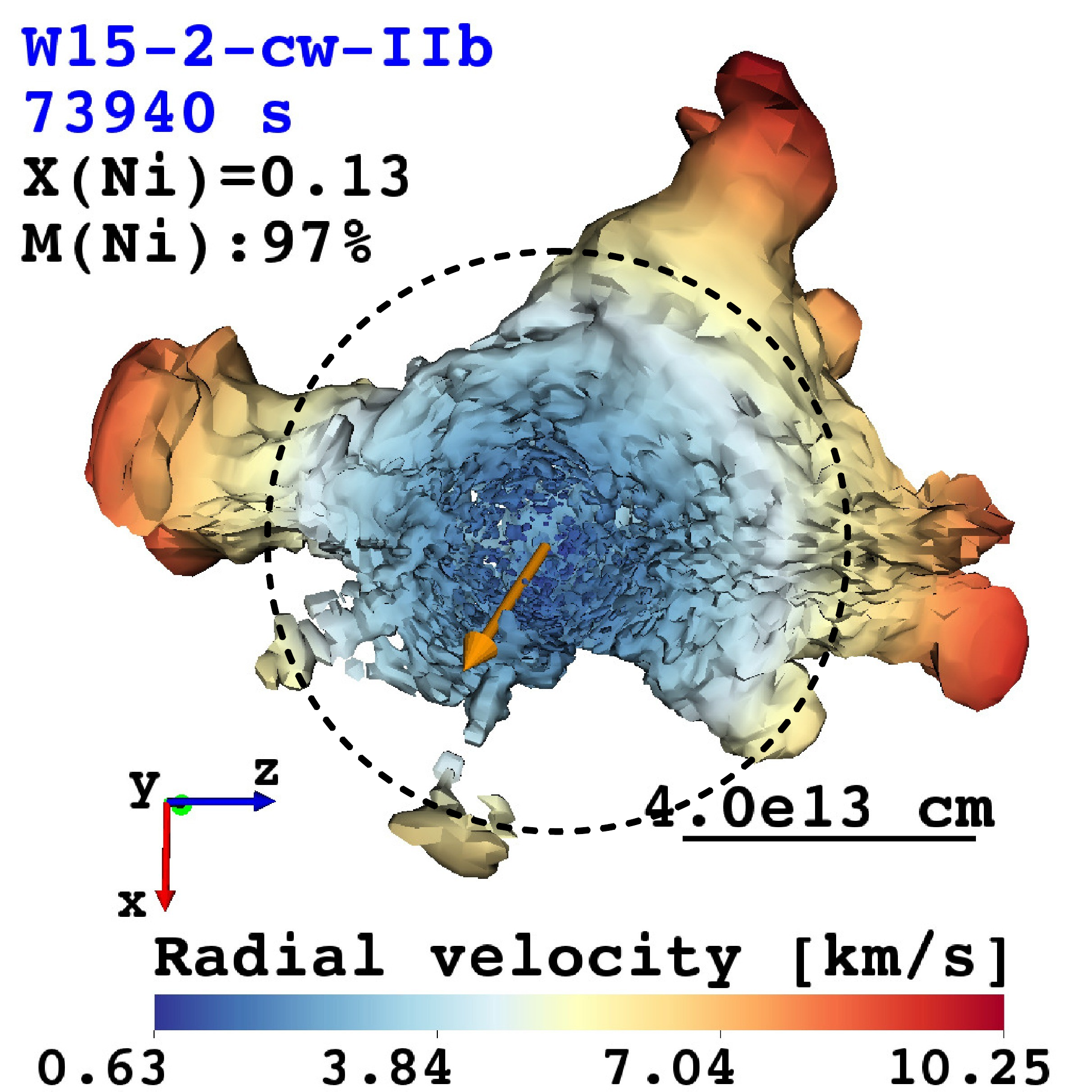}
 \caption{{\em Left:} Iron K-shell emission around the reverse-shock 
sphere of Cas~A as measured by the Chandra X-ray observatory
(\cite{DeLaney10,Hwang12}). The NS is marked by a white bullet.
(Image copied from 3D visualization available at 
{\tt http://3d.si.edu/explorer?modelid=45}.)
{\em Right:} Distribution of $^{56}$Fe in a 3D simulation of a 
neutrino-driven explosion (\cite{Wongwathanarat16}).
The NS kick direction (red arrow) and the approximate location of the
reverse shock (dashed circle, assuming 50\% of the iron to be shock 
heated) are indicated.} 
   \label{fig:CasAiron}
\end{center}
\end{figure}

{\underline{\it $^{44}$Ti and iron in Cassiopeia~A}}.
Using NuSTAR observations \cite{Grefenstette14} and
\cite{Grefenstette17}
mapped the 3D distribution of $^{44}$Ti in Cas~A, which is a 
perfect tracer of
the explosion geometry of this SN. They found a clumpy, 
non-uniform distribution of titanium around the centre of expansion
with most of the high-emission $^{44}$Ti knots being concentrated
in the hemisphere opposite to the NS kick direction.
Without any fine tuning, one of the 3D neutrino-driven explosion
simulations of \cite{Wongwathanarat15} is able to 
account for the main properties of the observed titanium and iron
distributions (Figs.~\ref{fig:CasAiron+ti} and \ref{fig:CasAiron};
\cite{Wongwathanarat16}). The theoretical model can explain 
the total yields of these nuclei estimated for Cas~A, their
3D geometry in relation to the NS kick magnitude and direction,
their radial and velocity distributions, and the relative variations
of the $^{44}$Ti/iron ratio suggested by the observations.
The morphology of this explosion seems to be compatible with
a neutrino-driven explosion that produced three large-scale plumes
of neutrino-heated, high-entropy ejecta that are essentially
located in one plane (\cite{Wongwathanarat16}), which could
be associated with the ``tilted thick disk'' identified in Cas~A
(see \cite{Grefenstette17}).

\section{Conclusions}

Neutrino-driven explosion simulations in 3D can explain basic
observational properties of the distribution of the innermost ejecta
(iron-group material, radioactive $^{44}$Ti) observed in young SN
remnants such as SN~1987A and Cas~A. Invoking new features in an
ad hoc manner (e.g.\ ``jittering jets'', \cite{Soker17a}) instead of
relying on self-consistently developing hydrodynamic instabilities 
that collaborate with the neutrino-heating mechanism, is not necessary.

For SN~1987A neutrino-driven explosions predict the existence of
a NS as a compact remnant. In view of a diagnosed explosion energy of
$(1.3-1.5)\times 10^{51}$\,erg, massive fallback is highly unlikely
for the progenitors (with He-core masses between 4 and 6\,$M_\odot$)
considered for SN~1987A. Because of a maximum NS mass of at least
2\,$M_\odot$ (\cite{Antoniadis13,Demorest10}) the collapse
of the compact remnant of a neutrino-driven explosion to a BH can
be ruled out. If the existence of a BH in SN~1987A can be proven,
radical revisions of our understanding of the SN mechanism would
therefore be inescapable (e.g., \cite{Blum16}).

\bigskip\bigskip\noindent
{\bf Acknowledgments.} This work
was supported by the Deutsche Forschungsgemeinschaft through
the Cluster of Excellence EXC 153 ``Origin and Structure of the Universe'', 
by the European Research Council through grant ERC-AdG No.\ 341157-COCO2CASA.
The computations were performed on Hydra of the Max Planck
Computing and Data Facility.

\end{document}